\documentclass[aps,prl, twocolumn,a4paper,showpacs,floatfix,superscriptaddress]{revtex4}
\usepackage{amssymb}
\usepackage{amsfonts}
\usepackage{amsmath,bm}
\usepackage{graphics}
\usepackage{color,times}
\usepackage{sidecap}
\usepackage{epsfig}

\begin{document}

\title{Contact-based Social Contagion in Multiplex Networks}

\author{Emanuele Cozzo}
\affiliation{Institute for Biocomputation and Physics of Complex Systems (BIFI), University of Zaragoza, Zaragoza 50009, Spain}
\affiliation{Department of Theoretical Physics, University of Zaragoza, Zaragoza 50009, Spain}

\author{Raquel A. Ba\~nos}

\affiliation{Institute for Biocomputation and Physics of Complex Systems (BIFI), University of Zaragoza, Zaragoza 50009, Spain}
\affiliation{Department of Theoretical Physics, University of Zaragoza, Zaragoza 50009, Spain}

\author{Sandro Meloni}
\affiliation{Institute for Biocomputation and Physics of Complex Systems (BIFI), University of Zaragoza, Zaragoza 50009, Spain}

\author{Yamir Moreno}
\affiliation{Institute for Biocomputation and Physics of Complex Systems (BIFI), University of Zaragoza, Zaragoza 50009, Spain}
\affiliation{Department of Theoretical Physics, University of Zaragoza, Zaragoza 50009, Spain}
\affiliation{Complex Networks and Systems Lagrange Lab, Institute for Scientific Interchange, Turin, Italy}

\date{\today}

\begin{abstract}

We develop a theoretical framework for the study of epidemic-like social contagion in large scale social systems. We consider the most general setting in which different communication platforms or categories form multiplex networks. Specifically, we propose a contact-based information spreading model, and show that the critical point of the multiplex system associated to the active phase is determined by the layer whose contact probability matrix has the largest eigenvalue. The framework is applied to a number of different situations, including a real multiplex system. Finally, we also show that when the system through which information is disseminating is inherently multiplex, working with the graph that results from the aggregation of the different layers is inaccurate.

\end{abstract}

\pacs{89.75.Hc,89.20.-a,89.75.Kd}
 
\maketitle

Social contagion processes such as the adoption of a belief, the propagation of opinions and behaviors, and the massive social movements that have recently unfolded worldwide \cite{centola2010, young2009,valente96,rogers2003,christakis2007,borge2011,gonzalez2011} are determined by many factors, among which the structure of the underlying topology and the dynamics of information spreading \cite{alexv2012}. The advent of new communication platforms such as online social networks (OSN), has made the study of social contagion more challenging. Today, individuals are increasingly exposed to many diverse sources of information, all of which they value differently \cite{friedkin2011}, giving raise to new communication patterns that directly impact both the dynamics of information spreading and the structure of the social networks \cite{onnela2007,karsai2011,iribarren2011,miritello}. Admittedly, the commonplace multi-channel information spreading that characterizes the way we exchange information nowadays has not been studied so far. One way to address the latter is to consider that the process of contagion occurs in a system made up of different layers, i.e., in a multiplex network \cite{mucha2010,szell, lee, nicosia, baxter,barrett,cozzo,bianconi,gomez2013,granell}. Although many studies have dealt with social contagion and information spreading on social networks, they all consider the case in which transmission occurs along the contacts of a simplex, i.e., single-layer, system. Here we aim at filling this existing gap.

The dynamics of this kind of processes can be modeled using different classes of approaches. Threshold models \cite{granovetter78,watts2002,melnik2013,yagan2012a,yagan2012b,borge2013} assume that individuals enroll in the process being modeled if a given intrinsic propensity level, the threshold, is surpassed. Although this class of models is useful to address the emergence of collective behavior, they are generally designed to simulate a single contagion process and therefore individuals, once they are active, remain so forever. This is not convenient in many situations that are characterized by self-sustained activity patterns \cite{borge2011,gonzalez2011}. For instance, think of an online social network in which tags are used to identify the topic of the information being transmitted (like {\em hashtags} in Twitter): individuals can use the same tag many times, but they can also decide not to use it after a number of times, thus being again susceptible to the contagion or in the language of threshold models, inactive. The latter features can be captured if one uses epidemic-like models of social contagion \cite{rapoport,goffman,hill2013}. In particular, the Susceptible-Infected-Susceptible (SIS) model \cite{Murray}, a classical approach to the study of disease spreading, allows individuals to cyclically change their dynamical state from susceptible (i.e., exposed to the tag) to infected (actively participating in the spreading process) and back to susceptible. 

In this paper, we propose a contact-based Markov chain approach \cite{gomez2010} to study epidemic-like social contagion in multiplex networks. We derive the conditions under which the dynamics reaches a steady state with active (infected) individuals coexisting with non-adopters. Our results show that the dynamics of the multiplex system is characterized by a critical point that depends solely on the layer with the largest eigenvalue of the contact probability matrix. We also show how our modeling framework can be applied to different scenarios and that working with the network resulting from the projection of all layers (the aggregated network) is not accurate. 

\begin{figure}
\includegraphics[width=\columnwidth]{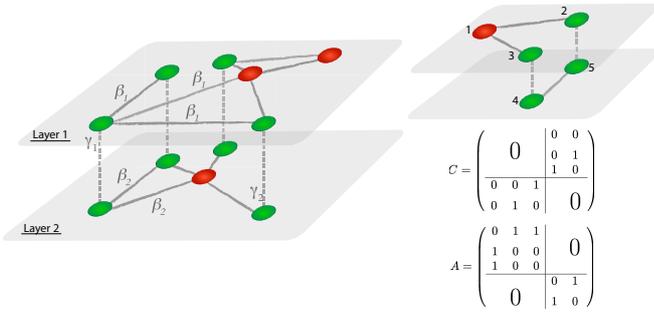}
\caption{(color online) Schematic of a 2-layer multiplex system where the contagion dynamics takes place. There are actors that take part in more than one layer (green nodes connected by the dotted edges), whereas others are present only in one layer (red nodes). 
$\beta_{1,2}$ is the contagion rate within the same layer whereas $\gamma_{1,2}$ represents the probability that the contagion occurs between layers. The right panel shows a small network and its associated $C$ and $A=\bigoplus_{\alpha}A_{\alpha}$.}
\label{fig1}
\end{figure}

Let us consider a multiplex system made up of $N$ nodes and $M$ layers (see Figure\ \ref{fig1}), and let the supra-contact probability matrix $\bar{R}=\{R_{ij}\}$ be
\begin{equation}
\bar{R}=\bigoplus_{\alpha} R_{\alpha}+\left(\vec{\frac{\gamma}{\beta}}\right)^{T}C
\end{equation}
where the $R_{\alpha}$'s are the contact probability matrices of each layer $\alpha$ and $C$ is the interlayer coupling matrix whose elements $C_{ij}=1$ if $i$ and $j$ represent the same actor in different layers. Thus it is a matrix with non-zero entries only in the off-diagonal blocks, see Fig.\ref{fig1}.  Moreover, for a given layer $\alpha$, $R_{\alpha}$ is defined as in the single-layer scenario \cite{gomez2010}, i.e., 
\begin{equation}
(R_\alpha)_{ij}=1-\left(1-\frac{(A_\alpha)_{ij}}{{k_\alpha}_i}\right)^{{\lambda_\alpha}_i},
\label{eqR}
\end{equation}
being $A_\alpha$ the adjacency matrix of layer $\alpha$ and $k_{{\alpha}_i}$ the degree of node $i$ in layer $\alpha$. In addition, all vectors are column vectors of the form $\vec{x}^T=(x_1\vec{1}_1^T,\dots,x_M\vec{1}_m^T)$, and $\vec{1}_\alpha$ are the vectors of all $1$s whose size is equal to the number of nodes $N_\alpha$ in layer $\alpha$. Thus, $\bar{R}$ is a block matrix with the $R_\alpha$ on the diagonal blocks and $\frac{\gamma_{l_i}}{\beta_{l_i}}C_{l_il_j}$ on the off-diagonal block $(l_i,l_j)$.
As in the simplex network, in each layer, the parameter ${\lambda_\alpha}_i$ determines the number of contacts that are made, so that one may go from a contact process (one contact per unit time) when ${\lambda_\alpha}_i=1$ to a fully reactive process (all neighbors within the layer are contacted) in the limit ${\lambda_\alpha}_i\longrightarrow\infty$ \cite{note1}. Moreover, the contagion between the layers is characterized by the ratio $\frac{\gamma_\alpha}{\beta_\alpha}$, where $\beta_\alpha$ is the rate at which the contagion spreads in layer $\alpha$. Finally, $\gamma_\alpha$ has the same meaning of $\beta$ but characterizes how contagion spreads from other layers to layer $\alpha$ (see Fig.\ \ref{fig1}), i.e., it is the rate at which a node in layer $\alpha$ gets infected if its counterparts in others layers are infected.

With the above ingredients, it is easy to see that the discrete-time evolution equation for the probability of contagion of a node $i$ of the multiplex system has the same functional form as in the single-layer case \cite{gomez2010}, namely, 
\begin{eqnarray}
\vec{p}(t+1)&=&(\vec{1}-\vec{p}(t))*(\vec{1}-\vec{q}(t))+(\vec{1}-\vec{\mu})*\vec{p}(t)\nonumber\\
&+&\vec{\mu}*(\vec{1}-\vec{q}(t))*\vec{p}(t),
\label{reaceq}
\end{eqnarray}
where $*$ stands for elements' wise multiplication of two vectors, i.e., $(\vec{p}*\vec{q})_i=p_iq_i$ and $\vec{\mu}$ is a vector whose components are the rates at which adopters are again susceptible. Moreover, $q_{i}(t)$ is the probability that node $i$ will not be infected by any neighbor
\begin{equation}
q_i(t)=\prod_{j}(1-\beta R_{ij}p_{j}(t)).
\label{eq:q}
\end{equation}

Let us now assume that $\frac{\gamma_\alpha}{\beta_\alpha}=\frac{\gamma}{\beta}$ \cite{note2} and $\frac{\mu_\alpha}{\beta_\alpha}=\frac{\mu}{\beta}, \forall \alpha=1,\dots,M$ \cite{note3}. The phase diagram can be studied by solving Eq.\ (\ref{reaceq}) at the stationary state
\begin{equation}
\vec{p}=(1-\vec{q})+(1-\vec{\mu})\vec{p}*\vec{q
}\label{fix}
\end{equation}
This equation has always the trivial solution $p_i = 0$, $\forall i = 1,\dots, N$. Other non-trivial solutions are given by non zero fixed points of Eq.\ (\ref{fix}) and can be easily computed numerically by iteration. Linearizing $q_i$ around $0$, at first order we get
\begin{equation}
[\bar{R}-\frac{\mu}{\beta}I]p=0
\label{autoval}
\end{equation}
that has non-trivial solutions if and only if $\frac{\mu}{\beta}$ is an eigenvalue of $\bar{R}$.  Since we are looking for the onset of the macroscopic social contagion, namely, the critical point, the lowest value of $\frac{\beta}{\mu}$ satisfying Eq.\ (\ref{autoval}) is
\begin{equation}
\left(\frac{\beta}{\mu}\right)_c=\frac{1}{\bar{\Lambda}_{max}},
\end{equation}
where $\bar{\Lambda}_{max}$ is the largest eigenvalue of the matrix $\bar{R}$.

It is worth analyzing this result by means of a perturbative analysis. Let $\bar{\Lambda}_{max}\simeq \Lambda + \epsilon\triangle \Lambda$, where $\Lambda$ is the largest eigenvalue of $R=\bigoplus_\alpha R_\alpha$ and consider $\bar{R}=R+\epsilon C$, with $\epsilon=\frac{\gamma}{\beta}\ll 1$. Since $R$ is a block diagonal matrix, it has the same set of eigenvalues of $\{R_\alpha\}$ and thus we can analyze the system in terms of the largest eigenvalues of the contact matrices $R_\alpha$ of the layers $\alpha$. For simplicity, we take the calculation in the case of two layers (i.e., $\alpha=1,2$), but generalization to any number of layers is straightforward. The change in the eigenvalue (eigenvector) can be estimated using a first order approximation \cite{milanese2010} 
\begin{equation}
\bigtriangleup \Lambda_{max}= \frac{\vec{v}^TC\vec{v}}{\vec{v}^T\vec{v}},
\end{equation}
\begin{equation}
\bigtriangleup \vec{v}= \frac{C}{\Lambda}\vec{v},
\end{equation}
where $\vec{v}$ is the eigenvector associated to the largest eigenvalue $\Lambda$ of the unperturbed matrix $R$.
Two cases are possible: {\em i)} $\Lambda_1\gg \Lambda_2$ ($\Lambda_2\gg \Lambda_1$ is completely equivalent), and {\em ii)} $\Lambda_1\simeq \Lambda_2$, where $\Lambda_1$ ($\Lambda_2$) is the largest eigenvalue of $R_1$ ($R_2$). In the first case, the eigenvector associated to the largest eigenvalue $\Lambda=\Lambda_1$ is
\begin{equation}
\vec{v}=\left(
\begin{array}{c}
\vec{v}_{(1)}  \\ 
0
\end{array}\right).
\label{bara}
\end{equation}
Hence, $\bigtriangleup \Lambda= 0$ and
\begin{equation}
\bigtriangleup \vec{v}= \left(
\begin{array}{c}
0  \\ 
\frac{\epsilon}{\Lambda} \vec{v}_{(1)}
\end{array}\right).
\end{equation}
Therefore, at first order approximation, we have that the largest eigenvalue of $\bar{R}$ is $\bar{\Lambda}_{max}=\max_\alpha\{\Lambda_{\alpha}\}$, and hence the emergence of a macroscopic steady state for the dynamics is determined by the layer with the largest eigenvalue. We call that layer the dominant layer. Besides, the probability of a node to catch the contagion at the critical point in a non-dominant layer is also specified by the probability of being infected in the dominant one.

In the second case {\em (ii)}, the eigenvector associated with the largest eigenvalue $\Lambda=\Lambda_1=\Lambda_2$ is
\begin{equation}
\vec{v}=\left(
\begin{array}{c}
\vec{v}_{(1)}  \\ 
\vec{v}_{(2)}
\end{array}\right),
\end{equation}
where $\vec{v}_{(1)}$ ($\vec{v}_{(2)}$) is the eigenvector associated to $\Lambda_1$ ($\Lambda_2$). 
Thus, at first order we have
\begin{equation}
\bigtriangleup \Lambda= \frac{\vec{v}_{(1)}C_{12}\vec{v}_{(2)}+
\vec{v}_{(2)}C_{21}\vec{v}_{(1)}}{\vec{v}^T_{(1)}\vec{v}_{(1)}+\vec{v}^T_{(2)}
\vec{v}_{(2)}},
\end{equation}
and
\begin{equation}
\bigtriangleup \vec{v}= \left(
\begin{array}{c}
\frac{\epsilon}{\Lambda} \vec{v}_{(2)}  \\ 
\frac{\epsilon}{\Lambda} \vec{v}_{(1)}
\end{array}\right).
\end{equation}
The previous expression indicates that in this scenario, the critical point is smaller and that the correction depends on the relation between the eigenvector centralities of the nodes in both layers. To further analyze the dynamical features of the contagion process, we numerically solve the system of equations given by Eqs.\ (\ref{eq:q}) and (\ref{fix}) for the different scenarios considered above. In the first case, when $\Lambda_{1}\gg \Lambda_{2}$, the dynamics of the multiplex system is completely dominated by the layer with the largest eigenvalue of $R_\alpha$. Thus, we expect that the contagion threshold coincides with the one of the dominant layer and no effect of the inter-layer diffusion parameter $\epsilon= \frac{\gamma}{\beta}$ near the threshold. 

\begin{figure}
\includegraphics[width=0.9\columnwidth]{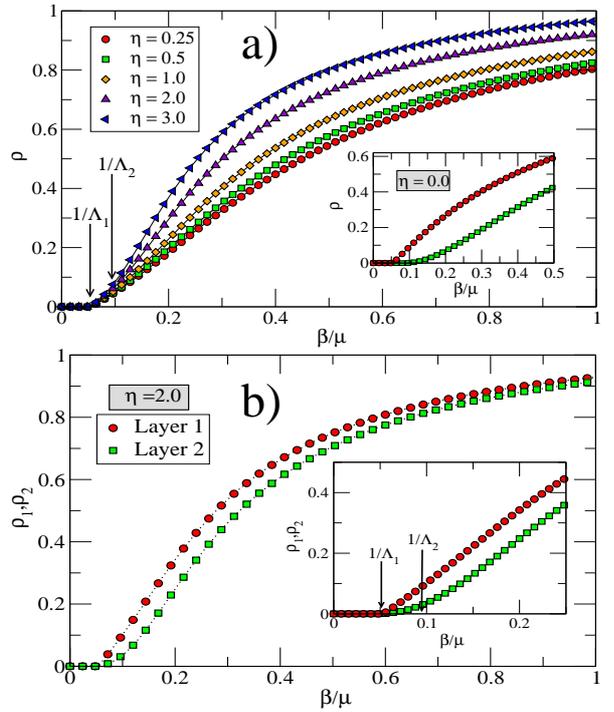}
\caption{(color online) Panel (a): Density of adopters ($\rho$) at the steady state against the rescaled contagion probability $\frac{\beta}{\mu}$ for a multiplex system composed of two layers with $N=10^4$ nodes each for different values of the ratio $\eta=\frac{\gamma}{\beta}$. The arrows represent the inverse of the largest eigenvalues of the two layers, whereas the inset shows the case in which both layers are completely disconnected. Panel (b): the same quantity of panel (a), for $\eta=2.0$,  is represented but computed at each layer. The inset is a zoom around the critical point. See the text for further details.}
\label{fig2}
\end{figure}

Figure \ref{fig2}a depicts the fraction of infectees, $\rho=\frac{1}{N}\sum_i p_i$, at the steady state against the rescaled contagion probability $\frac{\beta}{\mu}$ for a multiplex composed by two layers of $N_1=N_2=10^4$ nodes (thus $N=N_1+N_2=2\cdot10^4$). Both layers have been obtained using the uncorrelated configuration model with degree distribution $P(k) \sim k^{-g}$ with $g=2.3$ for the first layer and $g=3.0$ for the second one. Furthermore, we have assumed a fully reactive scenario in both layers of the system (i.e., $\lambda_1=\lambda_2\rightarrow\infty$ in Eq.\ (\ref{eqR})). As seen in panel (a), where arrows represent the inverse of the largest eigenvalues, the contagion threshold is set by $1/\Lambda_1$. It is worth noticing that the perturbative result still hold even for $\frac{\gamma}{\beta}=1$. This is due to the fact that the number of links added to the multiplex is small compared to the number of intra-layer links and the perturbation can still be considered small \cite{milanese2010}. On the other hand, the inset shows the results one would obtain if both layers were disconnected. In this case, each one would have their independent contagion thresholds determined by their largest eigenvalues.

It is also of interest to inspect the phase diagrams of the two layers separately. This is what is shown in Fig.\ \ref{fig2}b, where we represent the fraction of infectees at the steady state of each layer. As already discussed, the dominant layer fixes the contagion threshold of the multiplex network. However, it also induces a shift of the critical point of the second layer to smaller values. In other words, the multiplex nature of the system leads to an earlier transition to an active phase also in the non-dominant layer, as its critical point is now smaller than the expected value for the isolated system, i.e., $(\frac{\beta}{\mu})_{c_2}<\frac{1}{\Lambda_2}$. 

\begin{figure}
\includegraphics[width=\columnwidth]{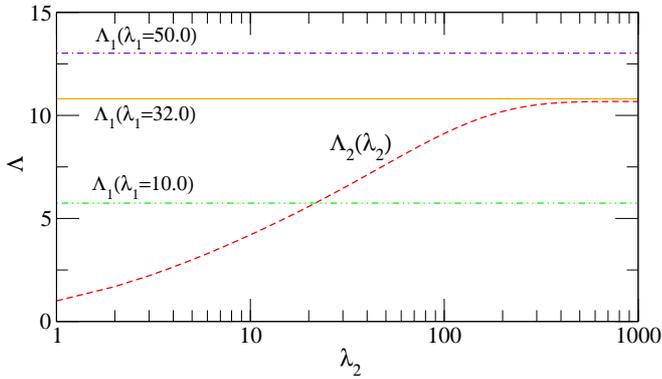}
\caption{(color online) Dependence of the largest eigenvalues of the contact probability matrices, $R_{\alpha}$'s, on $\lambda_{\alpha}$ for the system in Fig.\ \ref{fig2}. As it can be seen, there might be a crossover signaling that the dominant layer changes. This cross-over occurs only if the activity of the topologically dominant layer is small enough: in the example, it should be smaller than $\lambda_1=32$.}
\label{fig3}
\end{figure}

Furthermore, a unique feature of the model directly linked to the multiplex nature of the system is worth stressing. As the largest eigenvalues involved in the calculations are those associated to the matrices $R_{\alpha}$, they depend not only on the adjacency matrices $A_{\alpha}$, but also on $\lambda_{\alpha_{i}}$ (see Eq.\ (\ref{eqR})). This dependency has an interesting and novel effect as shown in Fig.\ \ref{fig3}: as the $\lambda_{\alpha}$'s characterize the number of effective contacts per unit time, a layer that does not prevail in the contagion dynamics because it is not topologically dominant (in terms of its $A_{\alpha}$) can compensate its lack of structural strength by increasing $\lambda_{\alpha}$ so as to eventually become the one with the largest eigenvalue of the multiplex network. The previous feature opens the door to potential applications in which by tuning the activity on one layer, the latter can take over the rest of the system and set its critical properties. Similarly, the above mechanism could explain situations in which the system is in the critical region despite the fact that by observing one layer one would expect the contrary. In other words, to determine whether the system is in a critical regime, one should have access to both the topological and activity features of all layers. This is in line with the findings in \cite{brummitt}, however, our model shows that once the dominant layer (if there is one) is detected, the analysis of the system dynamics can be carried out only on that layer.

We have also explored the scenario {\em ii)}, $\Lambda_1 \simeq \Lambda_2$, for which the largest eigenvalue of the multiplex is given as $\Lambda_{max} = \max_{\{1,2\}}\{\Lambda_{1},\Lambda_{2}\} + O(\epsilon)$. In particular, as one needs two networks with similar (very similar in this case) largest eigenvalues, we have used the same network in each layer and reshuffled the nodes from one layer to another to avoid correlation between the degree and the neighborhood of a node in the two layers. Also in this case (figure not shown), numerical results confirm the theoretical expectation. 

\begin{figure}
\includegraphics[width=\columnwidth]{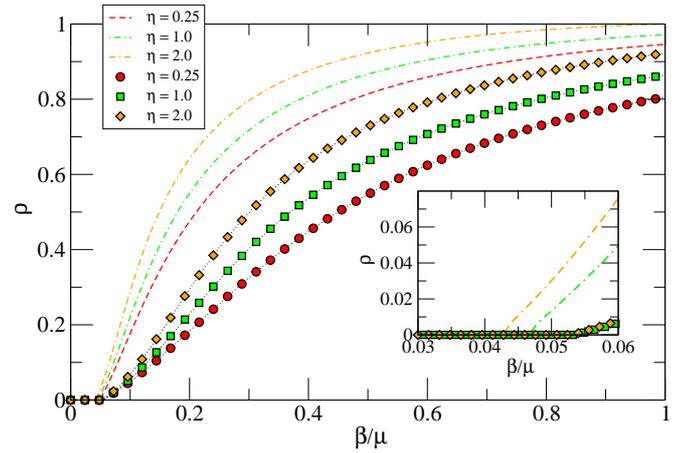}
\caption{(color online) Density of adopters ($\rho$) at the steady state as a function of the rescaled contagion probability $\frac{\beta}{\mu}$ for a multiplex system composed of two layers with $N=10^4$ nodes each (lines with symbols) and the corresponding aggregated graph (dotted lines). Different curves represent different values of the ratio $\eta=\frac{\gamma}{\beta}$ as indicated. The inset is a zoom of the region around the contagion threshold.}
\label{fig4}
\end{figure}

Next, we study the differences in the contagion process when considering the contraction along the inter-layer links of the multiplex. This amounts to consider an aggregated graph that corresponds to a simplex network in which all nodes and their respective links in each layer have been grouped together, and where the inter-layer connections between the same nodes are represented as self-loops. Since the largest eigenvalue of the contraction is larger than that of the multiplex, we expect the contagion threshold of the projected network to be smaller than that of the multiplex system. In addition, the number of infectees at the steady state should also be smaller for the multiplex network, since the correction to the probabilities of being infected, $p_i$'s, is small in this system.  Figure \ref{fig4} shows results of numerical calculations for both systems. As it can be seen more clearly in the inset of panel, the contagion thresholds are different. More importantly, the figure provides grounded evidences of why one cannot reduce a system that is inherently multi-level to a projected network $-$ the observed level of prevalence significantly differs from one system to the other. For instance, fixing the ratio $\frac{\beta}{\mu}$ that characterizes the spreading process within one layer, one can get estimates for the contagion incidence as higher as twice the actual value (that of the multiplex network).

In summary, we have proposed a contact-based framework to study the dynamics of social contagion processes in multiplex networks. Several results are worth highlighting. First, we have shown that the contagion threshold of the multiplex system is determined by the largest eigenvalue of the contact probability matrices of the layers that made up the system. Second, when a layer is dominant, the transition to a global steady state is driven by the dynamics at that layer. In this situation, the coupling between layers also affects the critical properties of the non-dominant layers by lowering their contagion thresholds. Furthermore, we have convincingly shown that disregarding the inherent multiplex nature of a system by dealing with the corresponding aggregated graph could lead to wrong conclusions. Our results could help understanding the spreading of information in multilevel socio-technical systems and how users behavior (via either $\gamma_{\alpha}$ or $\lambda_{{\alpha}_i}$) might modify the critical properties of contagion processes. Finally, our analyses suggest that there are three different ways in which the ``competitiveness'' (as far as its potential for contagion is concerned) of a layer can be enhanced: increasing the size of the layer, the connectivity of its nodes or their activity.

\begin{acknowledgments}

E. C and R.A.B. were supported by the FPI program of the Government of Arag\'on, Spain. This work has been partially supported by MINECO through Grants FIS2011-25167 and FIS2012-35719; Comunidad de Arag\'on (Spain) through a grant to the group FENOL and by the EC FET-Proactive Projects PLEXMATH (grant 317614, to YM) and MULTIPLEX (grant 317532 to YM and SM).

\end{acknowledgments}


\begin{thebibliography}{}
\bibitem{note1} In this limit $\bar{R}$ reduces to the supra-adjacency matrix $\bar{A}=A+C$ except for the fact that $\bar{A}$ is defined with $C$ and not with $\bm{\frac{\gamma}{\beta}}C$. In other words, $\bar{R}$ reduces to $\bm{\frac{\gamma}{\beta}}*\bar{A}$, where $\bm{\frac{\gamma}{\beta}}$ is a matrix with all $1$s on the diagonal blocks and $\frac{\gamma}{\beta}I$ on the off-diagonal blocks and $*$ denotes the Hadamard product.
\bibitem{note2} This does not change the results below as fas as $\frac{\gamma_{\alpha}}{\beta_{\alpha}}$ be small $\forall \alpha$.
\bibitem{note3} It can be shown that our main conclusions hold also when the ratio $\frac{\mu_{\alpha}}{\beta_{\alpha}}$ varies across different layers. However, assuming that $\frac{\mu_\alpha}{\beta_\alpha}=\frac{\mu}{\beta}, \forall \alpha=1,\dots,M$ implies that what remains roughly the same across layers is the ratio between $\mu$ and $\beta$. This seems suitable since a higher contagion rate can be reasonably associated with a higher recovery rate, because the 'pathogen' is the same and then the only factor influencing $\beta$ and $\mu$ is the easiness with which opinion/behavior changes (which is set by the context).
\end{thebibliography}
\end{document}